\documentclass[seceq]{ptptex}
\usepackage{wrapft}

\newcommand {\beq}{\begin{equation}}
\newcommand {\eeq}{\end{equation}}
\newcommand {\bea}{\begin{eqnarray}}
\newcommand {\eea}{\end{eqnarray}}
\newcommand {\nn}{\nonumber}
\newcommand {\tr}{{\rm tr\,}}

\newcommand {\dd}{\mbox{d}}

\newcommand {\del}{\partial}

\newcommand {\bA}{\tt A}
\newcommand {\bB}{\tt B}
\newcommand {\bC}{\tt C}

\title{
A Lattice Formulation of Super Yang-Mills Theories with Exact Supersymmetry\footnote{
Report number: OIQP-04-1\\
This presentation is based on the works~\citen{sugino,sugino2,sugino3}.}
}

\author{
Fumihiko \textsc{Sugino}%
}

\inst{
Okayama Institute for Quantum Physics, Kyoyama 1-9-1, Okayama 700-0015, Japan
}

\abst{
We construct SU($N$) super Yang-Mills theories 
with extended supersymmetry on hypercubic lattices of various dimensions 
keeping one or two supercharges exactly. 

It is based on topological field theory formulation for the super Yang-Mills 
theories. 
Gauge fields are represented by compact unitary link variables, 
and the exact supercharges on the lattice are nilpotent up to gauge transformations.  
In particular, the lattice models are free from the vacuum degeneracy problem, which 
was encountered in earlier approaches. 
Thus, we do not need to introduce any supersymmetry breaking terms, 
and the exact supersymmetry is preserved wholly in the process of taking the continuum limit.  

Among the models, we show that the 
desired continuum theories are obtained without any fine tuning of parameters 
for the cases ${\cal N}=2, 4, 8$ in two-dimensions. 
Also, the cases ${\cal N}=4, 8$ in three-dimensions are investigated, 
and a problem arising in four-dimensional models is discussed. 
}

\begin{document}

\maketitle

\section{Introduction}
Nonperturbative aspects 
in supersymmetric gauge theory are quite interesting not only from the 
field-theoretical point of view beyond the standard model, 
but also from the AdS/CFT duality between gauge theory and gravity in 
string theory.\cite{maldacena} 

A conventional approach to the nonperturbative study is lattice formulation, 
which enables numerical analysis for any 
observables not restricted to special operators with 
the BPS saturated or chiral properties. 
However, there has been difficulty on the lattice approach to supersymmetry, 
because of lack of infinitesimal translational invariance on the lattice
and breakdown of the Leibniz rule \cite{fujikawa, elitzur}. 
In spite of the difficulty, it is possible to 
construct lattice models, which do not have manifest supersymmetry but 
flow to the desired supersymmetric theories in the continuum limit. 
One of the examples is ${\cal N}=1$ super Yang-Mills (SYM) theory in 
four-dimensions whose field contents are gauge bosons and gauginos. 
Since in the theory 
the only relevant supersymmetry breaking operator is the gaugino mass, 
one can arrive at the supersymmetric 
continuum theory if the radiative corrections are 
not allowed to induce the relevant operator by symmetries realized  
in the lattice theory. Making use of domain wall or overlap fermions 
keeps discrete chiral symmetry on lattice, which is the symmetry excluding 
the fermion mass \cite{nishimura}.    

Supersymmetric theories with extended supersymmetry have some 
supercharges, which are not related to the infinitesimal translations 
and can be seen as fermionic internal symmetries. It is possible to realize 
a part of such supercharges as exact symmetry on lattice, and 
the exact supersymmetry is expected to play a key role to restore the full 
supersymmetry in the continuum limit with fine tuning of a few or no parameters. 

For SYM theories with extended supersymmetry, 
Cohen, Kaplan, Katz and \"{U}nsal recently proposed 
such a kind of various lattice models motivated by the idea of 
deconstruction.\cite{kaplan, kaplan2}
\footnote{
For some related works, see Refs.~\citen{giedt}. 
Also, for other attempts to lattice formulations of supersymmetry, 
see Refs.~\citen{kikukawa-nakayama,other,catterall,kawamoto}.} 
In these models, to generate the kinetic terms of the target theories 
and to stabilize noncompact bosonic zero-modes (the so-called radions), 
one has to add terms softly breaking the exact supersymmetry, 
which are tuned to vanish in large volume limit.   

In many cases, the above `internal' supersymmetries  
can be reinterpreted as the BRST symmetries in topological field theories. 
So, in order to construct lattice models respecting the `internal' 
supersymmetry, it seems natural to start with the topological field theory 
formulation of the theory. In Ref.~\citen{catterall}, 
Catterall discussed on a general formulation 
of lattice models based on the connection to topological field theory 
for supersymmetric theories without gauge symmetry.\footnote{
For related works, see Refs.~\citen{kawamoto}.}   

Here, standing on the same philosophy, 
we construct lattice models for SYM theories with extended 
supersymmetry keeping one or two supercharges exactly. 
Our models are motivated by the topological field theory formulation of 
${\cal N}=2, 4$ SYM theories, 
and free from the radion problems. 
The lattices have hypercubic structures, 
and the gauge fields are expressed as ordinary 
compact unitary variables on the lattice links. 

In sections 2 and 3, we construct lattice models for two-dimensional ${\cal N}=2, 4$ 
SYM theories based on (balanced) topological field theory formulation, and discuss on 
renormalization near the continuum limit. 
In sections 4 and 5, starting naive lattice actions for four-dimensional ${\cal N}=2, 4$ 
SYM theories, we construct lattice models for ${\cal N}=4, 8$ in two-dimensions and for 
${\cal N}=8$ in three-dimensions. 
Section 6 is devoted to the summary and discussion on some future directions. 
   
Throughout this paper, 
we focus on the gauge group $G = {\rm SU}(N)$. 
At the points discussing continuum 
theories, notations of repeated indices in formulas are assumed to be summed. 
On the other hand, 
when treating 
lattice theories, we explicitly write the summation over the indices 
except the cases of no possible confusion.  

\setcounter{equation}{0}
\section{2D ${\cal N}=2$ SYM}

\subsection{Continuum Action}
The action of ${\cal N}=2$ SYM in two-dimensions 
can be written as the 
`topological field theory (TFT) form' \cite{witten}: 
\beq
S_{2D{\cal N}=2} = Q \frac{1}{2g^2}\int \dd^2x \, \tr \left[
\frac14 \eta\, [\phi, \,\bar{\phi}] -i\chi\Phi
+\chi H -i\psi_{\mu}D_{\mu}\bar{\phi}\right], 
\label{TFT_S}
\eeq
where $\mu$ is the index for two-dimensional space-time. 
Bosonic fields are gauge fields $A_{\mu}$, complex scalars $\phi$, 
$\bar{\phi}$, and auxiliary field $H$. 
The other fields $\psi_{\mu}$, $\chi$, $\eta$ are fermionic,  
and $\Phi=2F_{12}$. 
$Q$ is one of the supercharges of ${\cal N}=2$ supersymmetry, 
and its transformation rule is given as   
\bea
QA_{\mu} = \psi_{\mu}, & & \quad Q\psi_{\mu} = iD_{\mu}\phi, \nn \\
Q\phi = 0, & &    \nn \\
Q\chi = H, & & \quad QH = [\phi, \,\chi], \nn \\
Q\bar{\phi} = \eta, & & \quad Q\eta = [\phi, \,\bar{\phi}]. 
\label{Q_continuum}
\eea
$Q$ is nilpotent up to infinitesimal gauge transformations 
with the parameter $\phi$. 
Note that the action has U$(1)_R$ symmetry whose charge assignment is  
$+2$ for $\phi$, 
$-2$ for $\bar{\phi}$, $+1$ for $\psi_{\mu}$, 
$-1$ for $\chi$ and $\eta$, $0$ for $A_{\mu}$ and $H$.
\subsection{Lattice Supersymmetry $Q$}
We formulate the theory (\ref{TFT_S}) 
on the two-dimensional square lattice 
keeping the supersymmetry $Q$. 
In the lattice theory, gauge fields $A_{\mu}(x)$ 
are promoted to the compact unitary variables 
\beq
U_{\mu}(x)= e^{iaA_{\mu}(x)}
\label{unitary}
\eeq 
on the link $(x, x+\hat{\mu})$. 
`$a$' stands for the lattice spacing, and $x\in {\bf Z}^2$ the lattice site. 
All other variables are distributed at sites.  
Interestingly, the $Q$-transformation (\ref{Q_continuum}) is extendible 
to the lattice variables preserving the property 
\beq
Q^2 = (\mbox{infinitesimal gauge transformation with the parameter } \phi)
\label{Q_nilpotent}
\eeq
as follows: 
\bea
 & & QU_{\mu}(x) = i\psi_{\mu}(x) U_{\mu}(x), \nn \\
 & & Q\psi_{\mu}(x) = i\psi_{\mu}(x)\psi_{\mu}(x) 
    -i\left(\phi(x) - U_{\mu}(x)\phi(x+\hat{\mu})U_{\mu}(x)^{\dagger}\right),
  \nn \\
 & & Q\phi(x) = 0,     \nn \\
 & & Q\chi(x) = H(x), \quad 
           QH(x) = [\phi(x), \,\chi(x)], \nn \\
 & & Q\bar{\phi}(x) = \eta(x), \quad  Q\eta(x) = [\phi(x), \,\bar{\phi}(x)]. 
\label{Q_lattice}
\eea
Also, 
$QU_{\mu}(x)^\dagger = -iU_{\mu}(x)^\dagger \psi_{\mu}(x)$ follows from 
$U_{\mu}(x)U_{\mu}(x)^\dagger =1$. 
All transformations except $QU_{\mu}(x)$ and 
$Q\psi_{\mu}(x)$ are of the same form as in the continuum case. 
Since (\ref{Q_nilpotent}) means  
\bea
Q^2U_{\mu}(x) & = & i(Q\psi_{\mu}(x))\,U_{\mu}(x)
-i\psi_{\mu}(x)\,(QU_{\mu}(x)) 
\nn \\
 & = & \phi(x)U_{\mu}(x)-U_{\mu}(x)\phi(x+\hat{\mu}),    
\eea
if we assume the formula ``$QU_{\mu}(x) = \cdots$'',
the transformation $Q\psi_{\mu}(x)$ is determined\footnote{The first term 
in the RHS does not vanish because 
$i\psi_{\mu}(x)\psi_{\mu}(x)= 
-\frac{1}{2}\sum_{a,b,c}f^{abc}\psi_{\mu}^a(x)\psi_{\mu}^b(x)T^c$ with $f^{abc}$ being 
structure constants of the gauge group.}. 
Then, happily  
$Q^2\psi_{\mu}(x)=[\phi(x), \,\psi_{\mu}(x)]$ is satisfied, 
and the $Q$-transformation is consistently closed. 
Note that we use the dimensionless variables here, and that 
various quantities are of the following orders:  
\bea
& & \psi_{\mu}(x), \chi(x), \eta(x) = O(a^{3/2}), \quad 
\phi(x), \bar{\phi}(x) = O(a), \quad H(x) = O(a^2), \nn \\
 & & Q=O(a^{1/2}). 
\label{order_of_a}
\eea
The first term in the RHS of ``$Q\psi_{\mu}(x)=\cdots$'' in (\ref{Q_lattice}) 
is of 
subleading order $O(a^3)$ and irrelevant in the continuum limit. 

\subsection{Lattice Action}
Once we have the $Q$-transformation rule closed among lattice variables, 
it is almost straightforward to construct the lattice action with the exact 
supersymmetry $Q$: 
\bea
S^{{\rm LAT}}_{2D{\cal N}=2} & = & Q\frac{1}{2g_0^2}\sum_x \, \tr\left[ 
\frac14 \eta(x)\, [\phi(x), \,\bar{\phi}(x)] -i\chi(x)(\Phi(x) +\Delta\Phi(x))
+\chi(x)H(x)\right. \nn \\
 & & \hspace{2cm}\left. \frac{}{} 
+i\sum_{\mu=1}^2\psi_{\mu}(x)\left(\bar{\phi}(x) - 
U_{\mu}(x)\bar{\phi}(x+\hat{\mu})U_{\mu}(x)^{\dagger}\right)\right], 
\label{lat_N=2_S}
\eea
where 
\bea
\Phi(x) & = & -i\left[U_{12}(x)- U_{21}(x)\right], 
\label{Phi_2d} \\
\Delta\Phi(x) & = & -r(2-U_{12}(x)-U_{21}(x)). 
\label{DeltaPhi_2d}
\eea
$U_{\mu\nu}$ are plaquette variables written as
\beq
U_{\mu\nu}(x) \equiv U_{\mu}(x) U_{\nu}(x+\hat{\mu}) 
U_{\mu}(x+\hat{\nu})^{\dagger} U_{\nu}(x)^{\dagger}. 
\eeq
The action (\ref{lat_N=2_S}) is clearly $Q$-invariant 
from its $Q$-exact form, and is U$(1)_R$ symmetric. 
It is an almost straightforward latticization of the continuum 
action (\ref{TFT_S}) except the terms containing $\Delta\Phi(x)$. 
We will explain a role of $\Delta\Phi(x)$. 

After acting $Q$ in the RHS, the action takes the form 
\bea
S^{{\rm LAT}}_{2D{\cal N}=2} & = & \frac{1}{2g_0^2}\sum_x \, \tr\left[
\frac14 [\phi(x), \,\bar{\phi}(x)]^2 + H(x)^2 
-iH(x)(\Phi(x) + \Delta\Phi(x)) \right. \nn \\
 & & 
+\sum_{\mu=1}^2\left(\phi(x)-U_{\mu}(x)\phi(x+\hat{\mu})U_{\mu}(x)^{\dagger}
\right)\left(\bar{\phi}(x)-U_{\mu}(x)\bar{\phi}(x+\hat{\mu})
U_{\mu}(x)^{\dagger}\right) \nn \\
 & & -\frac14 \eta(x)[\phi(x), \,\eta(x)] 
- \chi(x)[\phi(x), \,\chi(x)] \nn \\
 & & 
-\sum_{\mu=1}^2\psi_{\mu}(x)\psi_{\mu}(x)\left(\bar{\phi}(x)  + 
U_{\mu}(x)\bar{\phi}(x+\hat{\mu})U_{\mu}(x)^{\dagger}\right) 
+ i\chi(x)Q(\Phi(x)+\Delta\Phi(x)) \nn \\
 & & \left. \frac{}{}
-i\sum_{\mu=1}^2\psi_{\mu}(x)\left(\eta(x)-
U_{\mu}(x)\eta(x+\hat{\mu})U_{\mu}(x)^{\dagger}\right)\right]. 
\label{lat_N=2_S2}
\eea
In order to see the relevance of $\Delta\Phi(x)$, let us consider the case without 
$\Delta\Phi(x)$ in the action. 
After integrating out $H(x)$, induced $\Phi(x)^2$ term   
yields the gauge kinetic term as the form 
\beq
\frac{1}{2g_0^2}\sum_x
\sum_{\mu < \nu}\tr\left[-(U_{\mu\nu}(x) - U_{\nu\mu}(x))^2\right], 
\label{gauge_kin}
\eeq
which is different from the standard Wilson action 
\beq
\frac{1}{2g_0^2}\sum_x\sum_{\mu < \nu}\tr\left[2-U_{\mu\nu}(x)-U_{\nu\mu}(x)
\right]. 
\label{standard_wilson}
\eeq
In contrast with (\ref{standard_wilson}) giving 
the unique minimum $U_{\mu\nu}(x)=1$, 
the action (\ref{gauge_kin}) has many classical vacua  
\beq
U_{\mu\nu}(x) = \mbox{diag } (\pm 1, \cdots, \pm 1) 
\label{huge_minima}
\eeq
up to gauge transformations, where any combinations of $\pm 1$ 
with `$-1$' appearing even times are allowed in the diagonal entries. 
Since the configurations (\ref{huge_minima}) can be 
taken freely for each plaquette, it leads a huge degeneracy of  
vacua with the number growing as exponential of the number of the plaquettes. 
In order to see the dynamics of the model, we need to sum up contributions 
from all the minima,  
and the ordinary weak field 
expansion around a single vacuum 
$U_{\mu\nu}(x)=1$ can not be justified.\footnote{This kind of difficulty was 
already pointed out in Ref.~\citen{elitzur}.}  
Thus, we can not say anything about the continuum limit of the lattice 
model (\ref{lat_N=2_S2}) without its nonperturbative investigations.  
In order to resolve the difficulty without affecting the $Q$-supersymmetry, 
we introduce the $\Delta \Phi(x)$ terms with an appropriate choice of 
the parameter $r=\cot\theta$:\footnote{For a discussion about how the degeneracy 
is removed, see Ref.~\citen{sugino2}.}
\beq
e^{i2\ell\theta}\neq 1 \quad \mbox{for } \forall \ell=1, \cdots, N. 
\eeq

\subsection{About Fermion Doublers}
We expand the exponential of the link variable (\ref{unitary}), 
and look at the kinetic terms in the action (\ref{lat_N=2_S2}).   
Because in the bosonic sector no species doublers appear, 
in the fermionic sector also no doublers are expected 
due to the exact supersymmetry $Q$ of (\ref{lat_N=2_S2}). 
Let us see the fermionic sector explicitly. 

After rescaling each fermion variable by $a^{3/2}$ as indicated in  
(\ref{order_of_a}), the fermion kinetic terms are expressed as 
\beq
S_f^{(2)} =  \frac{a^4}{2g_0^2}\sum_{x, \mu}\tr\left[
-\frac12\Psi(x)^T\gamma_{\mu}(\Delta_{\mu}+\Delta^*_{\mu})\Psi(x) 
-a\frac12\Psi(x)^TP_{\mu}\Delta_{\mu}\Delta^*_{\mu}\Psi(x)\right], 
\label{wilson_like_N=2}
\eeq
where fermions were combined as 
$\Psi^T = \left( \psi_1, \psi_2, \chi, \frac12\eta \right)$. 
The $\gamma$-matrices and $P_{\mu}$ are given by 
\beq
\gamma_1=-i\sigma_1\otimes\sigma_1, \quad 
\gamma_2=i\sigma_1\otimes \sigma_3, \quad
P_1 = \sigma_1\otimes\sigma_2, \quad 
P_2 = \sigma_2\otimes{\bf 1}_2 
\eeq
with $\sigma_i$ ($i=1,2,3$) being Pauli matrices. 
Note that they all anticommute each other: 
\beq
\{\gamma_{\mu}, \gamma_{\nu}\} = -2\delta_{\mu\nu}, \quad 
\{ P_{\mu}, P_{\nu}\} = 2\delta_{\mu\nu}, \quad 
\{ \gamma_{\mu}, P_{\nu}\} = 0. 
\label{anticommute_N=2}
\eeq
$\Delta_{\mu}$ and $\Delta^*_{\mu}$ represent forward and backward difference  
operators respectively: 
\beq
\Delta_{\mu}f(x) \equiv \frac{1}{a}\left(f(x+\hat{\mu})-f(x)\right), \quad 
\Delta^*_{\mu}f(x) \equiv \frac{1}{a}\left(f(x)-f(x-\hat{\mu})\right).
\eeq
The kernel of the kinetic terms (\ref{wilson_like_N=2}) is written in the 
momentum space $-\frac{\pi}{a}\leq q_{\mu} < \frac{\pi}{a}$ as 
\beq
D=\sum_{\mu=1}^2\left[-i\gamma_{\mu}\frac{1}{a}\sin \left(q_{\mu}a\right)
+2P_{\mu}\frac{1}{a}\sin^2 \left(\frac{q_{\mu}a}{2}\right)\right]. 
\eeq
It is easy to see that the kernel $D$ vanishes only at the origin 
$q_1=q_2=0$, because using (\ref{anticommute_N=2}) we get 
\beq
D^2 = \frac{1}{a^2}\sum_{\mu=1}^2\left[\sin^2\left(q_{\mu}a\right)
+4 \sin^4\left(\frac{q_{\mu}a}{2}\right)\right]. 
\label{D_square}
\eeq
Thus, the fermion kinetic terms contain no fermion doublers. 
The last term containing $P_{\mu}$ in (\ref{wilson_like_N=2}) 
has a similar structure to the Wilson 
term, and plays a role of removing fermion doublers. 
Since the lattice action is U$(1)_R$ invariant and the fermion doublers are 
removed keeping the chiral U$(1)_R$, the model must break some 
assumptions of Nielsen-Ninomiya's no go theorem \cite{nielsen-ninomiya}.    
In fact, broken is 
the assumption ``{\it There exists a conserved charge $Q_F$ 
corresponding to the fermion number.}''.  
When combining fermions into a two-component Dirac spinor as 
\beq
\zeta = \frac{1}{\sqrt{2}}
\left(\begin{array}{c}\psi_1 -i\psi_2 \\
                      \chi + i\frac12\eta \end{array}\right), \quad 
\bar{\zeta} = \frac{1}{\sqrt{2}}
\left(\psi_1+i\psi_2, \, \chi -i \frac12\eta\right), 
\eeq
$Q_F$ corresponds to the U$(1)_J$ rotation: 
$
\zeta \rightarrow e^{i\theta}\zeta, \quad 
\bar{\zeta} \rightarrow e^{-i\theta} \bar{\zeta}. 
$
The first term in (\ref{wilson_like_N=2}), 
giving a naive fermion kinetic term on the lattice, is written as  
the combination $\bar{\zeta}_{\alpha}\zeta_{\beta}$, which is invariant 
under the U$(1)_J$. On the other hand, the last term containing $P_{\mu}$ 
takes the form: 
\beq
\frac{a^4}{2g_0^2}\sum_x\frac{a}{2}\, \tr\left[ \varepsilon_{\alpha\beta} 
\zeta_{\alpha}(\Delta_1\Delta^*_1 +i\Delta_2\Delta^*_2)\zeta_{\beta} - 
\varepsilon_{\alpha\beta} 
\bar{\zeta}_{\alpha}(\Delta_1\Delta^*_1 -i\Delta_2\Delta^*_2)
\bar{\zeta}_{\beta} \right] 
\eeq
to break the U$(1)_J$ invariance. 

\subsection{Renormalization}
At the classical level, the lattice action (\ref{lat_N=2_S}) leads to 
the continuum action (\ref{TFT_S}) in the limit $a\rightarrow 0$ 
with $g^{-2} \equiv a^{2}g_0^{-2}$ kept fixed, 
and thus the ${\cal N}=2$ supersymmetry and rotational symmetry 
in two-dimensions are restored. 
We will check whether the symmetry restoration persists against 
quantum corrections, i.e. whether symmetries of the lattice action 
forbid any relevant or marginal operators induced which 
possibly obstruct the symmetry restoration.  
%
\begin{wraptable}{c}{\halftext}
\begin{tabular}{|c|ccccc|}
\hline \hline
$p=a+b+3c$ & 
\multicolumn{5}{|c|}{$\varphi^a\del^b\psi^{2c}$}  \\ 
\hline
0 &     &          &  1       &       &    \\
1 &     &          & $\varphi$  &      &     \\
2 &     &          & $\varphi^2$ &     &     \\
3 &     &$\varphi^3$, & $\psi\psi$, & $\varphi\del\varphi$  &     \\
4 & $\varphi^4$, & $\varphi^2\del\varphi$, & $(\del\varphi)^2$, & 
   $\psi\del\psi$, & $\varphi\psi\psi$ \\
\hline \hline
\end{tabular}
  \caption{List of operators with $p\leq 4$. 
}
\label{tab:operators}
\end{wraptable}
%

%
Assuming that the model has the critical point $g_0=0$ from the asymptotic freedom, 
we shall consider the renormalization effect perturbatively. 
The mass dimension of the coupling $g^2$ is two. 
For generic boson field $\varphi$ (other than the auxiliary fields) 
and fermion field $\psi$, the dimensions 
are 1 and 3/2 respectively.   
Thus, operators of the type $\varphi^a \del^b\psi^{2c}$ 
have the dimension $p\equiv a+b+3c$, where 
`$\del$' means a derivative with respect to the coordinates. 
{}From dimensional analysis, 
we can see that the operators receive the following 
radiative corrections up to some powers of possible logarithmic factors: 
\beq
\left(\frac{a^{p-4}}{g^2} + c_1 a^{p-2} + c_2 a^pg^2 + \cdots\right)
\int \dd^2x \, \varphi^a \del^b \psi^{2c},  
\label{hosei_2d}
\eeq
where $c_1, c_2, \cdots$ are constants dependent on $N$. 
The first, second and third terms in the parentheses 
represent the contributions at 
tree, one-loop 
and two-loop levels. It is easily seen from the fact that $g^2$ appears 
as an overall factor in front of the action and plays the same role as 
the Planck constant $\hbar$. 
Due to the super-renormalizable property of two-dimensional theory, 
the relevant corrections terminate at the two-loop. 
{}From the above formula, it is seen that 
the following operators can be relevant or marginal  
in the $a\rightarrow 0$ limit: operators with 
$p\leq 2$ induced at the one-loop level and with $p=0$ at the two-loop level. 
Operators with $p\leq 4$ are listed in Table \ref{tab:operators}. 

Since the identity operator does not affect the spectrum,    
we have to check operators of the types $\varphi$ and $\varphi^2$ only. 
Gauge symmetry and U$(1)_R$ invariance\footnote{Note that the U$(1)_R$ symmetry is not 
anomalous for $G={\rm SU}(N)$ in the two-dimensions. } 
allow the operator 
$\tr \phi\bar{\phi}$, while it is forbidden by 
the supersymmetry $Q$.  
Hence,  
no relevant or marginal operators except the identity  
are generated by radiative corrections, which means that     
in the continuum limit 
full supersymmetry and rotational symmetry are 
considered to be restored without any fine tuning. 

\setcounter{equation}{0}
\section{2D ${\cal N}=4$ SYM}

\subsection{Continuum Action}
The action of ${\cal N}=4$ SYM in two-dimensions can be written as the following 
`Balanced Topological Field Theory (BTFT) form':\cite{vafa-witten,dijkgraaf-moore}  
\bea
S_{2D{\cal N}=4} & = &  Q_+Q_-{\cal F}_{2D{\cal N}=4}, \nn \\
{\cal F}_{2D{\cal N}=4} & = & \frac{1}{2g^2}\int  \dd^2x \, \tr\left[
-iB\Phi - \psi_{+\mu}\psi_{-\mu} - \chi_+\chi_- -\frac14\eta_+\eta_-
\right], 
\label{BTFT_S_2d}
\eea
where $Q_{\pm}$ are two of supercharges of the ${\cal N}=4$ theory, 
and $\Phi \equiv 2F_{12}$. Bosons are 
gauge fields $A_{\mu}$ ($\mu=1,\,2$) and 
scalar fields $B$, $C$, $\phi$, $\bar{\phi}$. 
Also, there are auxiliary fields $\tilde{H}_{\mu}$, $H$. 
Other fields $\psi_{\pm\mu}$, $\chi_{\pm}$, $\eta_{\pm}$ are 
fermions. Transformation rule of the supersymmetry $Q_{\pm}$ is given by 
\bea
& & Q_+A_{\mu}= \psi_{+\mu}, \quad Q_+\psi_{+\mu} = iD_{\mu}\phi, \quad 
Q_-\psi_{+\mu} = \frac{i}{2}D_{\mu}C-\tilde{H}_{\mu}, \nn \\
& & Q_-A_{\mu}= \psi_{-\mu}, \quad Q_-\psi_{-\mu} = -iD_{\mu}\bar{\phi}, \quad 
Q_+\psi_{-\mu} = \frac{i}{2}D_{\mu}C+\tilde{H}_{\mu}, \nn \\
 & & Q_+\tilde{H}_{\mu} = [\phi, \,\psi_{-\mu}]
-\frac12[C, \,\psi_{+\mu}] -\frac{i}{2}D_{\mu}\eta_+, \nn \\
 & & Q_-\tilde{H}_{\mu} = [\bar{\phi}, \,\psi_{+\mu}]
+\frac12[C, \,\psi_{-\mu}] +\frac{i}{2}D_{\mu}\eta_-, 
\label{group_A}
\eea
\bea
 & & Q_+B = \chi_+, \quad Q_+\chi_+ = [\phi, \,B], \quad 
Q_-\chi_+ = \frac12[C, \,B]-H, \nn \\
 & & Q_-B = \chi_-, \quad Q_-\chi_- = -[\bar{\phi}, \,B], \quad 
Q_+\chi_- = \frac12[C, \,B]+H, \nn \\
 & & Q_+H = [\phi, \,\chi_-] +\frac12[B, \,\eta_+] 
-\frac12[C, \,\chi_+],  \nn \\
 & &  Q_-H = [\bar{\phi}, \,\chi_+] -\frac12[B, \,\eta_-] 
+\frac12[C, \,\chi_-], 
\label{group_B_2d}
\eea
\bea
 & & Q_+C = \eta_+, \quad Q_+\eta_+ = [\phi, \,C], \quad 
Q_-\eta_+ = -[\phi, \,\bar{\phi}], \nn \\
 & & Q_-C = \eta_-, \quad Q_-\eta_- = -[\bar{\phi}, \,C], \quad 
Q_+\eta_- = [\phi, \,\bar{\phi}], \nn \\
 & & Q_+\phi = 0, \quad Q_-\phi= -\eta_+, \quad 
Q_+\bar{\phi} = \eta_-, \quad Q_-\bar{\phi} = 0. 
\label{group_C}
\eea
This transformation leads the following nilpotency of $Q_{\pm}$ 
(up to gauge transformations): 
\bea
Q_+^2 & = & 
(\mbox{infinitesimal gauge transformation with the parameter }\phi), \nn \\
Q_-^2 & = & 
(\mbox{infinitesimal gauge transformation with the parameter }-\bar{\phi}), 
\nn \\
\{Q_+, Q_-\} & = & 
 (\mbox{infinitesimal gauge transformation with the parameter }C). 
\label{nilpotent_N=4}
\eea

In this formulation, among the SU(4) internal symmetry of the ${\cal N}=4$ theory, 
its subgroup SU$(2)_R$ is manifest which rotates $(Q_+, Q_-)$. 
Under the SU$(2)_R$, each of $(\psi^a_{+\mu}, \psi^a_{-\mu})$, 
$(\chi^a_+, \chi^a_-)$, 
$(\eta^a_+, -\eta^a_-)$ and $(Q_+, Q_-)$ transforms as a doublet, 
and $(\phi^a, C^a, -\bar{\phi}^a)$ as a triplet. 
Also, let us note a symmetry of the action (\ref{BTFT_S_2d}) 
under exchanging the two supercharges $Q_+ \leftrightarrow Q_-$  
with 
\bea
 & & \phi \rightarrow -\bar{\phi}, \quad \bar{\phi} \rightarrow -\phi, \quad 
B \rightarrow -B, \nn \\
 & & \chi_+ \rightarrow -\chi_-, \quad \chi_- \rightarrow -\chi_+, \quad 
\tilde{H}_{\mu} \rightarrow -\tilde{H}_{\mu}, \nn \\
 & & \psi_{\pm\mu} \rightarrow \psi_{\mp\mu}, \quad 
\eta_{\pm} \rightarrow \eta_{\mp}. 
\label{Q_exchange_2d}
\eea
 
\subsection{Lattice Supersymmetry $Q_\pm$} 
Similarly to the ${\cal N}=2$ cases, it is possible to define the theory 
(\ref{BTFT_S_2d}) 
on the square lattice preserving the two supercharges $Q_{\pm}$. 
The transformation rule 
(\ref{group_A}) is modified as 
\bea
 & & Q_+U_{\mu}(x) = i\psi_{+\mu}(x)U_{\mu}(x), \nn \\
 & & Q_-U_{\mu}(x) = i\psi_{-\mu}(x)U_{\mu}(x), \nn \\
 & & Q_+\psi_{+\mu}(x) = i\psi_{+\mu}\psi_{+\mu}(x) 
  -i\left(\phi(x)-U_{\mu}(x)\phi(x+\hat{\mu})U_{\mu}(x)^{\dagger}\right), 
\nn \\
 & & Q_-\psi_{-\mu}(x) = i\psi_{-\mu}\psi_{-\mu}(x) 
  +i\left(\bar{\phi}(x)-U_{\mu}(x)\bar{\phi}(x+\hat{\mu})U_{\mu}(x)^{\dagger}
\right), \nn \\
 & & Q_-\psi_{+\mu}(x) = \frac{i}{2}
\left\{\psi_{+\mu}(x), \,\psi_{-\mu}(x)\right\} -\frac{i}{2}
\left(C(x)-U_{\mu}(x)C(x+\hat{\mu})U_{\mu}(x)^{\dagger}\right) 
-\tilde{H}_{\mu}(x), \nn \\
 & & Q_+\psi_{-\mu}(x) = \frac{i}{2}
\left\{\psi_{+\mu}(x), \,\psi_{-\mu}(x)\right\} -\frac{i}{2}
\left(C(x)-U_{\mu}(x)C(x+\hat{\mu})U_{\mu}(x)^{\dagger}\right) 
+\tilde{H}_{\mu}(x), \nn \\
 & & Q_+\tilde{H}_{\mu}(x) = -\frac12
\left[\psi_{-\mu}(x), \,\phi(x)+U_{\mu}(x)\phi(x+\hat{\mu})U_{\mu}(x)^{\dagger}
\right] \nn \\
 & & \hspace{2cm} 
+\frac14\left[\psi_{+\mu}(x), \, C(x) +U_{\mu}(x)C(x+\hat{\mu})
U_{\mu}(x)^{\dagger}\right] \nn \\
 & & \hspace{2cm} +\frac{i}{2}\left(\eta_+(x) 
-U_{\mu}(x)\eta_+(x+\hat{\mu})U_{\mu}(x)^{\dagger}\right) \nn \\
 & & \hspace{2cm}  
+\frac{i}{2}\left[\psi_{+\mu}(x), \,\tilde{H}_{\mu}(x)\right] 
+\frac14\left[\psi_{+\mu}(x)\psi_{+\mu}(x), \,\psi_{-\mu}(x)\right], \nn \\
 & & Q_-\tilde{H}_{\mu}(x) = -\frac12
\left[\psi_{+\mu}(x), \,\bar{\phi}(x)+U_{\mu}(x)\bar{\phi}(x+\hat{\mu})
U_{\mu}(x)^{\dagger}\right] \nn \\
 & & \hspace{2cm}
-\frac14\left[\psi_{-\mu}(x), \, C(x) +U_{\mu}(x)C(x+\hat{\mu})
U_{\mu}(x)^{\dagger}\right] \nn \\
 & & \hspace{2cm} -\frac{i}{2}\left(\eta_-(x) 
-U_{\mu}(x)\eta_-(x+\hat{\mu})U_{\mu}(x)^{\dagger}\right) \nn \\
 & & \hspace{2cm}
+\frac{i}{2}\left[\psi_{-\mu}(x), \,\tilde{H}_{\mu}(x)\right] 
-\frac14\left[\psi_{-\mu}(x)\psi_{-\mu}(x), \,\psi_{+\mu}(x)\right]. 
\label{group_A_lat}
\eea
The other transformations (\ref{group_B_2d}, \ref{group_C}) 
do not change the form under the 
latticization. Note that this modification keeps the nilpotency 
(\ref{nilpotent_N=4}). 
  
Making use of the $Q_{\pm}$-transformation rule 
in terms of lattice variables, we  
construct lattice actions with the exact supercharges $Q_{\pm}$ as 
\bea
S^{{\rm LAT}}_{2D{\cal N}=4} &  = &
Q_+Q_-\frac{1}{2g_0^2}\sum_x\, \tr \left[-iB(x)(\Phi(x) +\Delta\Phi(x))- 
\sum_{\mu=1}^2\psi_{+\mu}(x)\psi_{-\mu}(x) \right. \nn \\ 
 & & \hspace{2.5cm} \left. -\chi_+(x)\chi_-(x)-\frac14\eta_+(x)\eta_-(x)\right], 
\label{lat_N=4_S_2d}
\eea
where $\Phi(x)$ and $\Delta\Phi(x)$ are given by 
(\ref{Phi_2d}) and (\ref{DeltaPhi_2d}), respectively. 
Note that the lattice formulation retains the symmetries under SU$(2)_R$ 
as well as the $Q_+\leftrightarrow Q_-$ exchange. 

Similarly to the ${\cal N}=2$ case, $\Delta\Phi(x)$ removes the vacuum degeneracy and 
the fermion doublers do not appear.  
With respect to the renormalization argument, symmetries of the 
lattice action are sufficient to restore full supersymmetry and 
rotational invariance in the continuum limit. 
For instance, 
gauge invariance and   
SU$(2)_R$ symmetry allow the operators $\tr(4\phi\bar{\phi} + C^2)$ and 
$\tr B^2$, 
but they are not admissible from the supersymmetry $Q_{\pm}$. 
Thus, radiative corrections are not allowed to 
generate any relevant or marginal operators 
except the identity, which means the restoration of full supersymmetry and rotational 
invariance in the continuum limit.

\setcounter{equation}{0}
\section{3D ${\cal N}=4$}

Also for ${\cal N}=2$ theory in four-dimensions, 
we can write the action in the `TFT form', 
and construct a naive lattice action as 
\bea
S^{{\rm LAT}}_{4D{\cal N}=2} & = & Q\frac{1}{2g_0^2}\sum_x \, \tr\left[ 
\frac14 \eta(x)\, [\phi(x), \,\bar{\phi}(x)] 
-i\vec{\chi}(x)\cdot(\vec{\Phi}(x) + \Delta\vec{\Phi}(x))
+\vec{\chi}(x)\cdot\vec{H}(x)\right. \nn \\
 & & \hspace{2cm}\left. \frac{}{} 
+i\sum_{\mu=1}^4\psi_{\mu}(x)\left(\bar{\phi}(x) - 
U_{\mu}(x)\bar{\phi}(x+\hat{\mu})U_{\mu}(x)^{\dagger}\right)\right], 
\label{lat_4DN=2_S}
\eea
where $\vec{H}(x)$, $\vec{\chi}(x)$, $\vec{\Phi}(x)$ and $\Delta\vec{\Phi}(x)$ 
are three-component vectors, and 
\bea
\Phi_{\bA}(x) & = & -i\left[U_{4, -{\bA}}(x) -U_{-{\bA}, 4}(x) 
+ \frac12\sum_{\bB, \bC=1}^3\varepsilon_{\bA\bB\bC}\,
(U_{\bB\bC}(x) - U_{\bC\bB}(x))\right],   \\
\Delta\Phi_{1}(x) & = & -r\left[W_{4, -1}(x) +W_{23}(x)\right], \nn \\
\Delta\Phi_{2}(x) & = & -r\left[W_{4, -2}(x) +W_{31}(x)\right], \nn \\
\Delta\Phi_{3}(x) & = & -r\left[W_{4, -3}(x) +W_{12}(x)\right]. 
\eea
$W_{\mu\nu}(x)$ are defined by
\beq
W_{\mu\nu}(x)\equiv 2-U_{\mu\nu}(x)-U_{\nu\mu}(x), \quad 
U_{-\mu}(x) = U_{\mu}(x-\hat{\mu})^\dagger. 
\eeq
The vacuum degeneracy is removed with 
the choice $r=\cot\theta$: $0<\theta \leq \frac{\pi}{2N}$.\footnote{For 
a detailed discussion, see Ref.~\citen{sugino3}.} 
It turns out, however, that the quadratic terms in $A_{\mu}$ in  
$\tr (\vec{\Phi}(x) + \Delta\vec{\Phi}(x))^2$ 
have surplus zero-modes (other than gauge degrees of freedom)
carrying the nonzero momentum in the fourth direction: $q_4=\pm\frac{\pi}{2a}$. 
Fermion kinetic terms also have zero-modes at the same momenta, 
which is consistent to the exact supersymmetry $Q$ because the surplus modes are not 
exact zero-modes of the full action (only of the quadratic terms) 
and a fermionic partner necessarily exist for each 
bosonic surplus modes. 

Here, we do not resolve the problem, but 
consider the dimensional reduction with respect to the fourth direction. 
Then, the four-dimensional ${\cal N}=2$ theory reduces to three-dimensional ${\cal N}=4$ 
theory where the surplus modes are all killed. 
Thus, the dimensional reduced lattice model reproduces 
desired three-dimensional ${\cal N}=4$ theory in the classical continuum limit. 
The renormalization argument tells that 
necessary is fine-tuning of three parameters for the counter terms 
with the mass dimension three:
\[
Q\sum_{\mu=1}^3\tr(\psi\bar{\phi}), \quad 
Q\,\tr(\psi_4\bar{\phi}), \quad 
Q\sum_{{\bA}=1}^3\tr(\chi_{\bA}A_4)  
\]
in order to arrive at the desired continuum theory at the quantum level. 


\setcounter{equation}{0}
\section{3D ${\cal N}=8$ and 2D ${\cal N}=8$}

We construct a naive lattice action for four-dimensional ${\cal N}=4$ SYM as 
\bea
S^{{\rm LAT}}_{4D{\cal N}=4} &  = & 
Q_+Q_-\frac{1}{2g_0^2}\sum_x\, \tr \left[
-i\vec{B}(x)\cdot(\vec{\Phi}(x) + \Delta\vec{\Phi}(x)) \right. \nn \\
& & - \frac13\sum_{\bA, \bB, \bC =1}^3\varepsilon_{\bA\bB\bC}\,B_{\bA}(x)\,
[B_{\bB}(x), \,B_{\bC}(x)] 
- \sum_{\mu=1}^4\psi_{+\mu}(x)\psi_{-\mu}(x) \nn \\
 & & \left. -\vec{\chi}_+(x)\cdot\vec{\chi}_-(x) 
-\frac14\eta_+(x)\eta_-(x)\right].  
\eea    
Here, the same problem of the surplus modes occurs. 

Considering the dimensional reduction with respect to the fourth direction, we obtain 
a lattice model for three-dimensional ${\cal N}=8$ SYM 
which reproduces the desired theory in the classical continuum limit. 
Also, further reduction with respect to the third direction leads two-dimensional 
${\cal N}=8$ theory.  

The result of the renormalization argument is as follows. 
For the three-dimensional ${\cal N}=8$ model, 
required is the one parameter fine-tuning for the operator of the mass dimension three:
\[
Q_+Q_-\tr[(B_1+B_2+B_3)A_4],  
\]
while the two-dimensional model of ${\cal N}=8$ 
needs no fine-tuning of parameters. 

\setcounter{equation}{0}
\section{Summary and Discussions}
\label{sec:summary}
We have constructed varioius lattice models for SYM theories of 
${\cal N}=2, 4, 8$ in two-dimensions 
and of ${\cal N}=4, 8$ in three-dimensions, based on `(balanced) topological field theory 
form' of the theories. 
The formulation exactly realizes a part of the supersymmetry and employes compact link 
variables for the gauge fields on hypercubic lattice. 
{}From the renormalization argument, we have shown that the desired continuum theories are 
obtained by fine-tuning three and one parameters for the three-dimensional ${\cal N}=4$ and 
8 theories respectively, while the two-dimensional theories require no tunings. 

We have also seen that there exist surplus modes in four-dimensional naive lattice models for 
${\cal N}=2, 4$. It may be related to exact realization of the topological term $\tr F\wedge F$ 
on the lattice which needs a nonabelian extension of the solution for the U(1) 
case.\cite{luscher} 

Finally, we mention some future directions related to this work. 
\begin{itemize}
\item
The models for ${\cal N}=8$ in two- and three-dimensions constructed here could be used 
for nonperturbative investigation of matrix string models.\cite{DVVBS}  

\item
It is interesting to do actual simulative studies utilizing the formulation presented here. 
As a first step, it would be necessary to investigate the positivity of the fermion determinant. 

\item
It would be possible to couple the lattice SYM models to matter fields by 
referring topological QCD formulation.\cite{sako}
\end{itemize}

\section*{Acknowledgements}
The author would like to thank all the participants and the organizers 
of the lattice theory workshop at YITP, ``SUSY 2004" at Tsukuba, 
``Lattice 2004" at FermiLab and ``Summer Institute 2004" at Fuji-Yoshida 
for making productive atmosphere and stimulating discussions.

%

\end{document}